\documentclass{article}

    \PassOptionsToPackage{numbers, compress}{natbib}

    \usepackage[preprint]{neurips_2022}

\usepackage[utf8]{inputenc} 
\usepackage[T1]{fontenc}    
\usepackage{hyperref}       
\usepackage{url}            
\usepackage{booktabs}       
\usepackage{amsfonts}       
\usepackage{nicefrac}       
\usepackage{microtype}      
\usepackage{xcolor}         

\usepackage{graphicx}
\usepackage{amsmath}
\usepackage{amssymb}
\usepackage{bbm}
\usepackage{bm}
\usepackage{booktabs}
\usepackage{xcolor}
\usepackage{array}
\usepackage[labelfont=bf,font=small,skip=3pt]{caption}
\newcolumntype{C}[1]{>{\centering\arraybackslash}p{#1}}
\usepackage{wrapfig}
\usepackage{subcaption}

\title{A Survey of IMU Based Cross-Modal Transfer Learning in Human Activity Recognition}

\author{
  Abhi Kamboj \\
  University of Illinois, Urbana-Champaign \\
  \texttt{akamboj2}
  \And
  Minh Do \\
  University of Illinois, Urbana-Champaign \\
  \texttt{minhdo@illinois.edu} \\
}

\begin{document}

\maketitle

\begin{abstract}
Despite living in a multi-sensory world, most AI models are limited to textual and visual understanding of human motion and behavior. 
Inertial measurement sensors provide a signal for AI to understand motion, however, in practice they has been understudied due to numerous difficulties and the uniterpretability of the data to humans. 
In fact, full situational awareness of human motion could best be understood through a combination of sensors.
In this survey we investigate how knowledge can be transferred and utilized amongst modalities for Human Activity/Action Recognition (HAR), i.e. cross-modality transfer learning.
We motivate the importance and potential of IMU data and its applicability in cross-modality learning as well as the importance of studying the HAR problem. 
We categorize HAR related tasks by time and abstractness and then compare various types of multimodal HAR datasets.
We also distinguish and expound on many related but inconsistently used terms in the literature, such as transfer learning, domain adaptation, representation learning, sensor fusion, and multimodal learning, and describe how cross-modal learning fits with all these concepts. 
We then review the literature in IMU-based cross-modal transfer for HAR.
The two main approaches for cross-modal transfer are instance-based transfer, where instances of one modality are mapped to another (e.g. knowledge is transferred in the input space), or feature-based transfer, where the model relates the modalities in an intermediate latent space (e.g. knowledge is transferred in the feature space).
Finally, we discuss future research directions and applications in cross-modal HAR.

\end{abstract}

\section{Introduction}

Our interaction with computing has historically been centered around visual and textual modalities, despite living in a world rich with diverse sensory inputs. 
Smartwatches and smartphones have enabled the integration of inertial sensors that provide insights into fundamental human movements, such as walking or interacting with humans or objects, yet these sensors have the potential to provide much more information on human motion and behavior.
Moreover, the strategic utilization of motion data in conjunction with other modalities, such as video, has the potential to significantly influence various fields including health monitoring, surveillance, human-robot interaction, autonomous driving, and more.

In the realm of human action recognition (HAR), deep learning methodologies have focused predominantly on visual data \cite{ji20123d, simonyan2014two, lin2022frozen, wang2023masked}, or multimodal vision language models \cite{radford2021learning, wang2022internvideo, tong2022videomae, piergiovanni2023rethinking}. 
However, Inertial Measurement Units (IMUs), which typically provide 3-axis acceleration and 3-axis gyroscopic information, emerge as a strong candidate for understanding human motion and behavior. 
Unfortunately, IMUs remain underutilized within current machine-learning approaches due to numerous difficulties.

Apart from IMU sensors, many works explore HAR through various modalities, including depth, wifi signals, audio, skeletal data, electrocardiogram (ECG) sensors and more.
Cross-modal learning has emerged as a method to transfer knowledge across modalities which can allow for improved HAR performance in systems where one modality's training data is limited.
We hypothesize that IMU-based cross-modality learning presents a compelling avenue for achieving enhanced performance in many HAR related tasks.
To substantiate this claim, we conduct a systematic review of the existing literature in this field.


The organization of this paper is as follows. 
In Section \ref{sec::motivation}, we further describe our motivation for studying IMU data (Subsection \ref{sec::motivation::IMU}) and the HAR task (Subsection \ref{sec::motivation::HAR}).
Section \ref{sec::related_works} discusses related survey papers and distinguishes cross-modal learning from sensor fusion, transfer learning, and domain adaptation (Subsections \ref{sec::related_works::sensor_fusion}, \ref{sec::related_works::transfer_learning}, and
\ref{sec::related_works::domain_adaptation} respectively.
In Section \ref{sec::cross_modal_transfer}, we categorize cross-modal learning into instance-based transfer, discussed in \ref{sec::cross_modal_transfer::instance}, and and feature-based transfer discussed in \ref{sec::cross_modal_transfer::feature}.
Finally, in Sections \ref{sec::discussion} and \ref{sec::conclusion}, we discuss the implications of cross-modal transfer in multimodal learning and conclude respectively.

\section{Motivation}
\label{sec::motivation}


\subsection{Inertial Measurement Units}
\label{sec::motivation::IMU}

Many individuals are not cognizant of the prevalence of IMUs in their daily lives, as these sensors are seamlessly integrated into commonplace devices such as smartphones, smartwatches, tablets, and even earbuds \cite{mollyn2023imuposer}.
The ubiquity of these signals provides ample opportunities for machine learning research to understand human motion. 
While humans find it relatively straightforward to accurately label a person's activity based on video or textual descriptions, the task becomes much less intuitive when attempting to interpret IMU graphs for the same purpose. 
Nevertheless, from a theoretical perspective, machine learning algorithms may find it simpler to learn patterns from the lower-dimensional IMU data.
IMUs provide 6-channel time series data; in contrast, a video contains on the scale of a 3x224x224 stream of time series data.
The manifold hypothesis argues that for a given purpose, the true meaning of high-dimensional data lies on a lower-dimensional manifold.
IMU data might be this natural lower-dimensional manifold for human activities that can be extracted from high-dimensional videos.

IMU data exhibits numerous properties that makes it a strong candidate for analyzing human motion and behavior \cite{lara2012survey}.
They enhance privacy preservation due to their inherent uninterpretability.
Compact physical form factor, size, and cost make them ideal for products.
The portability and non-intrusiveness of IMUs further contribute to their appeal, allowing for integration into wearable devices and facilitating natural, unencumbered movement during data collection.
The low-dimensional data translates into reduced memory storage and bandwidth requirement when transmitting and processing the data.
The inherent insensitivity of IMUs to ambient lighting conditions or visual obstruction ensures robust performance in various environments. 
In addition, IMUs provide real-time insight into the precise dynamics of motion, capturing subtle movements that might be difficult for visual sensors to accurately discern.

Nonetheless, there are many difficulties inhibiting IMUs from being a more potent tool for capturing human motion and behavior in systems nowadays.
Although an IMU is physically small and cheap, the user has to be interacting with it or somehow in contact with it for it to measure their movement, whereas a visual sensor can be ambient, stationed away from the user.
In addition, with more data visual sensors provide rich contextual information such as appearance and environmental cues and thus may be able to discern more complex human actions or interactions with other objects in the environment.
Finally, although IMU data is more precise, it is only relative to one limb or one point on the human's body which may constrain how much information it can convey about the human's actions.
In addition, studies have shown that IMU data tends to "drift" with respect to its initial condition, making it a poor indicator of the IMU's relative location to objects in its vicinity or its own previous state \cite{dai2022hsc4d}.

Given the various strengths and weaknesses of various sensors, it is evident that  human motion understanding through a unimodal method can become more robust by incorporating information from multiple modalities.
Due to the advantageous attributes inherent in IMUs coupled with their underutilization and potential for success, the present literature review focuses on cross modal learning approachs for HAR that include IMU as a modality.


\subsection{Human Activity Recognition}
\label{sec::motivation::HAR}

Multimodal Human Activity Recogntion (HAR) with IMU technology has  profound impact on society across various domains including health, IoT and smart technology, robotics, graphics and business.

In the realm of healthcare, multimodal HAR plays a pivotal role in elderly care by enabling the detection of neurological disorders such as Parkinson's disease through gait and motion analysis.
It also facilitates the identification of mental health issues by assessing stress or depression based on a user's motion, thus contributing to early intervention and treatment.
Furthermore, the technology aids in simple health monitoring by detecign falls and notifying loved ones in case of emergencies. 
FOr individuals of all ages, biometric monitoring and exercise tracking promote healthy lifestyles and enhance self-awareness of personal health.
Additionally, the system can preemptively detect signs of medical emergencies such as strokes or heart attacks by analyzing users' movement and vital signs, potentially saving lives.

In the domain of IoT and smart technology multimodal HAR with IMU sensors enhances home comfort and energy optimization by understanding human motion and behavior patterns, contributing to more efficient energy usage and improving living conditions. 
It also strengthens security measures by identifying individuals through their motion, thereby enhancing home and city surveillance systems.
In urban settings, this technology aids in designing better traffic and crowd control systems by predicting human motion patterns.
Moreover, it enables efficient surveillance by identifying suspicious behavior and mitigating potential terrorist threats. 
Additionally, it assists in understanding energy usage for grid planning, thereby supporting the integration of renewable energy sources.

In the field of robotics, multimodal HAR with IMU sensors is instrumental in developing advanced prosthetics for healthcare applications and facilitating Human-Robot Interaction (HRI) in collaborative manufacturing and agricultural settings. It also contributes to the design of humanoid robots that can effectively navigate human environments and act as autonomous agents. Furthermore, the technology plays a crucial role in autonomous driving by predicting pedestrian intent and anticipating the behavior of other drivers on the road.

In the realm of graphics, multimodal HAR with IMU technology is pivotal for creating immersive Augmented Reality (AR) and Virtual Reality (VR) experiences, enabling realistic digital autonomous agents and simulations. These simulations are instrumental in training pilots, astronauts, and military personnel for real-world scenarios, as well as predicting human responses to apocalyptic events.

From a business perspective, multimodal HAR with IMU sensors offers valuable insights into consumer behavior, thereby informing marketing strategies and product development. It also contributes to workplace security and efficiency by monitoring workers' stress levels and health within office environments.

In conclusion, multimodal HAR with IMU technology has far-reaching impacts on society, revolutionizing healthcare, smart technology, robotics, graphics, and business practices. Its potential to improve human well-being, enhance technological capabilities, and inform decision-making processes underscores its significance as a transformative force in the modern world.







Firstly, we aim to delineate different levels of human motion understanding, providing a framework to shape our review and refine our terminology.
Many surveys exist discussing the defined human-related AI tasks such as human action detection \cite{sun2022human}, activity recognition problem \cite{jobanputra2019human, lara2012survey}, action prediction \cite{kong2022human}, activity localization \cite{bibbo2022overview} etc.
We  propose 2 methods to compare human related AI tasks, by time and by granularity. 

One metric to categorize human motion is by time in which the motion occurs.
Human or object detection/segmentation occurs in a single image. 
Human action recognition occurs over multiple frames or a few seconds.
Human activity recognition may occur over minutes or hours and long term human motion understanding may occur over years. 
Human behavior prediction attempts to understand the future, using any interval of the past.
The research landscape extensively addresses short-term Human Activity or Action Recognition (HAR) problems, whereas long-term human motion recognition and predication remain open and active areas of investigation. 
This is presumably attributed to the inherent challenges associated with collecting, processing, and conducting experiments over more extended time frames, underscoring the complexities inherent in prolonged temporal analyses.

\begin{figure}
    \centering
    \includegraphics{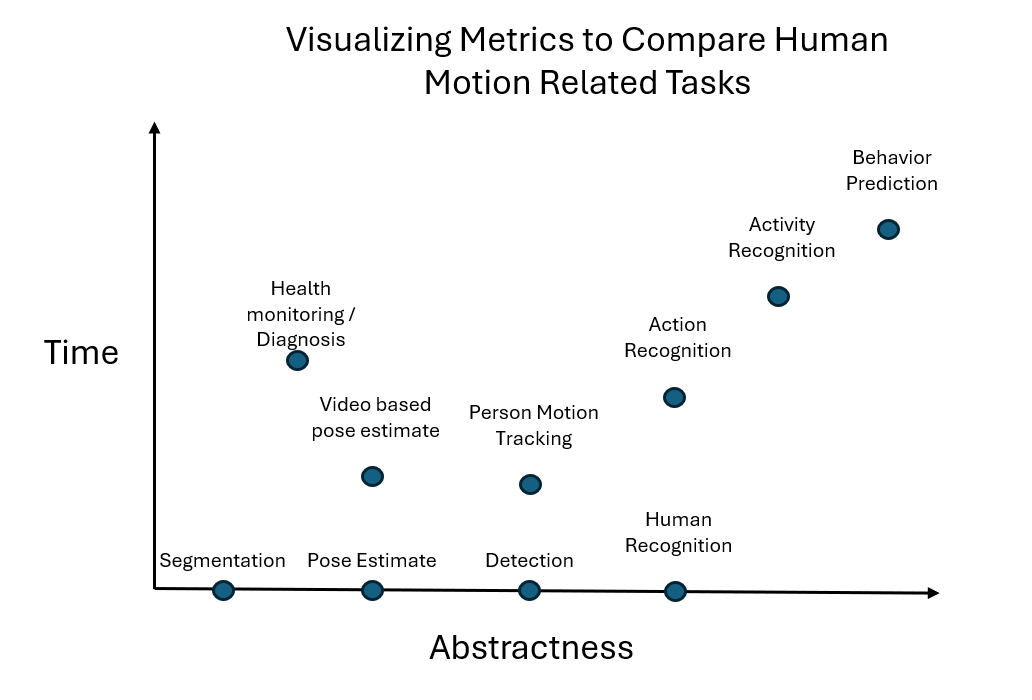}
    \caption{Displayed is a graph-like visualization plotting HAR-related tasks to compare each task with the other based on temporal length involved in the task as well as the abstractness or coarseness of the task.}
    \label{fig:har_metrics}
\end{figure}

Another metric to interpret human motion understanding involves the granularity or abstractness of the task.
Human pixel-wise segmentation is a very well defined precise task. 
Keypoint or pose estimation is a slightly more coarse task.
Human bounding box detection is even coarser.
Adding a temporal dimension yields action movement localization and action recognition.
A slightly more abstract problem is activity recognition.
At the highest level of abstraction we have motion or behavior understanding and prediction.
OpenPose \cite{cao2017realtime} represents a pivotal advancement in computer vision facilitating real-time estimation of human keypoints, underscoring the maturity of research in human pose estimation, which has trickled into human action recognition.
Nevertheless, the domain of activity recognition, along with more abstract tasks pertaining to human behavior, remains notably open and presents a compelling avenue for ongoing research exploration.
A visualization of the comparison amongst HAR related tasks is shown in Figure \ref{fig:har_metrics}.


In this spectrum of human action understanding, our review encompasses works pertaining to multimodal human action and activity recognition.
Action recognition usually involves identifying simple actions such as waving a hand, kicking a leg, or drawing a circle with your right hand.
This is mainly used for applications in HRI \cite{rodomagoulakis2016multimodal} and HCI \cite{sharma1998toward}.
Human activity recognition often involves one or more actions that have some semantic meaning together.
For example, playing soccer may include sub-activities of running and kicking a ball. 
Some works even explicitly use multiple action recognition steps to perform activity recognition \cite{li2020pastanet}.
One can further abstract actions and activities into higher categories. 
An abstraction of sub-activities to a larger activity may be referred to as a composite activity and multiple related activities occurring simultaneously are concurrent activities \cite{chen2021deep}.

There are numerous types of datasets for multimodal HAR applications.
Multimodal Human Action Datasets (MHAD) is a class of datasets that refer to short term action recognition and are commonly employed for gesture recognition and basic motion understanding in HRI, HCI and computer vision, e.g. UTD-MHAD \cite{chen2015utd}, USC-HAD \cite{zhang2012usc}, CZU-MHAD \cite{chao2022czu}. 
Activies of Daily Living (ADL) datasets focus on the slightly longer term task of activity recognition, and ADL datasets are often used for research on assisted living and smart home development, e.g. Toyota Smarthome \cite{das2019toyota} and MARBLE \cite{arrotta2021marble}. 
Egocentric multi-sensor datasets align IMU data with videos from the human perspective (e.g. a video from a GoPro) which are primarily used for AI in the applications of virtual/augmented reality or humanoid robotics, e.g. Ego4D \cite{grauman2022ego4d} and EpicKitchens \cite{damen2018scaling}, \cite{spriggs2009temporal}. 
Finally, autonomous driving datasets may also include IMU data and may be used for HAR related tasks (e.g. pedestrian segmentation, activity recognition or motion prediction), however, unless the dataset is measuring driver's interaction with the vehicle \cite{rosique2023autonomous}, the IMU data often pertains to the motion of the car as opposed to the motion of the human \cite{caesar2020nuscenes}. 

Our review looks at works that may span across one or more of these types of datasets as long as the work is focused on IMU based cross modal representation learning for HAR, where the IMU data pertains to human motion.
For example, a work that achieves state of the art performance on the UTD-MHAD dataset through a novel multimodal neural network and supervised training is not of interest to us.
A work that uses pretrained video representations to train an IMU action recognizer with less IMU data on the UTD-MHAD dataset is of interest to us, as it demonstrates cross-modal representation learning.



\section{Related Literature Reviews}
\label{sec::related_works}

\subsection{Sensor Fusion}
\label{sec::related_works::sensor_fusion}

Sensor fusion jointly uses multiple modalities to perform a task.
Cross-modal learning generally trains one modality first and then transfers the knowledge to the second.
Cross-modal learning may train a model using two modalities at once to learn the relation between them, however, only one modality is used at a time during inference resulting in the "crossing" of knowledge from one modality to another. 

For instance, suppose we have a camera and IMU and want to perform activity recognition.
A sensor fusion approach would collect and label video and IMU data and train a model to input video and IMU together and predict a label. 
The goal is to make the model predict more accurately because it can leverage the strengths and weaknesses of the other sensor. 
On the other hand, a cross-modal learning approach would first collect and label videos and train a model.
Then use the learned video representations to somehow transfer the knowledge to the IMU domain, resulting in a model that uses only the IMU data to perform HAR.
Both sensor fusion and cross-modal learning are considered multi-modal methods, however, cross-modal learning is a type of transfer learning (discussed in the next subsection) whereas sensor fusion is not.


Sensor fusion is often broken down into the following 3 categories based on where the data are combined \cite{ramachandram2017deep,majumder2020vision,sharma1998toward}: 
1) Early or data-level fusion combines the raw sensor outputs before any processing.
2) Middle/intermediate or feature-level fusion combines each sensor modality after some preprocessing or feature extraction.
3) Late or decision-level fusion combines the raw output, essentially ensembling separate models.
However, there do exist reviews that break fusion down into categories other than these\cite{chen2021deep}.

\begin{figure}
    \includegraphics[width=\columnwidth]{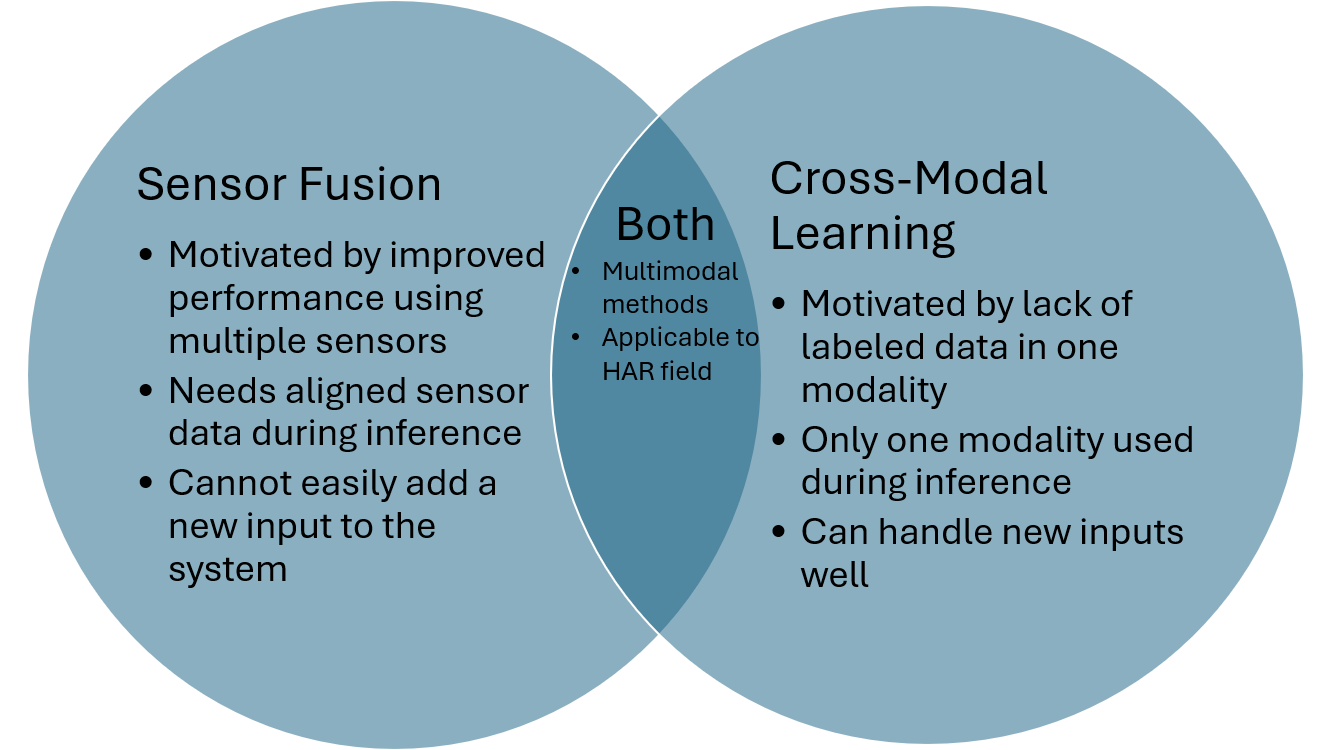}
    \hfill
    \caption{This diagram compares sensor fusion and cross-model transfer.}
    \label{fig:venn_digram}
\end{figure}

Cross-modal learning is more modular and practical in its use as it assumes one sensor has limited labeled data, whereas sensor fusion usually requires the data of all modalities, thus one can easily add and remove sensors to a cross-modal learning system.
On the other hand, sensor fusion may arguably perform better as it leverages the information in multiple sensors to enhance the performance of the system jointly. 
For a summary of comparisons, see Figure \ref{fig:venn_digram}.

There are have been many existing reviews on on sensor fusion in HAR.
The first works on HAR date back to the 90's, however, major advancements in performance came more recently with machine learning methods\cite{lara2012survey}. 
After machine learning, HAR started using deep learning models for more robust feature extraction for HAR in wearables \cite{attal2015physical, nweke2018deep, ramanujam2021human, zhang2022deep}. 
Other surveys broaden the focus by including works using various other types of sensors (e.g. ambient, object sensors, wearables, etc.) in their review\cite{wang2018deep, chen2021deep}.
Some works include or exclude only a certain subsets of sensors in their works such as IMU and vision \cite{majumder2020vision}
Ahad et al. \cite{ahad2021deep} further discusses how the transition from sensor-based HAR from handcrafted features to deep learning improved HAR performance, and highlights recent trends such as data-efficient active-learning \cite{hossain2017active} which queries a human annotator to label only the most meaningful points, as well as transferring models across modalities and environments.
The latter recent trend described in \cite{ahad2021deep} comprises the subject of this literature review, and no previous literature reviews focus on cross modal transfer in HAR. 



\subsection{Transfer Learning}
\label{sec::related_works::transfer_learning}
Assume we have some model $f:X^S->Y^S$ for some source distribution $S$.
In simple terms transfer learning attempts to learn some model $g:X^T->Y^T$ for some target distribution $T$ from the original model or data $f, X^S, Y^S$. 
For more robust/precise definitions, terminology and taxonomy of transfer learning refer to \cite{pan2009survey,weiss2016survey}.
Transfer learning tends to address insufficient labeled data, incompatible computation power or distributional mismatch between two machine learning tasks \cite{niu2020decade} and cross-modal learning similarily can address the same issues.

Continuing on the example of performing HAR with video and IMU data, we can demonstrate how cross-modal learning is a type of transfer learning.
As mentioned before, a cross-modal learning approach would first collect and label videos and train a model,  and then use the learned video representations to transfer knowledge to the IMU domain.
One method to do this could be learning how to predict IMU data from the video data or reusing/sharing the weights in the video model to create an IMU model to recognize activity.
These are examples of the two main methods of transfer learning instance transfer and feature-representation transfer respectively \cite{pan2009survey}.
In both cases, the goal for cross-modal learning is to adapt a model that uses one sensor as input data to use a different sensor's data as input, which may have less labeled data, i.e. have one sensor learn from the other's model.
The terminology used in different reviews varies when discussing transfer learning \cite{niu2020decade,weiss2016survey,pan2009survey}. 
Transfer learning can be broadly categorized into three settings: transductive transfer learning, inductive transfer learning, or unsupervised transfer learning \cite{pan2009survey}. 
This categorization depends on the availability of labels in the source and target domains, which is the main motivation behind transfer learning. 
Additionally, we consider whether the domains or tasks are the same or different in the source and target, as this will help us categorize cross-modal transfer.

Transductive learning involves using different domains, where one is labeled and the other lacks labels, but they correspond to the same task, $X^S \ne X^T, Y^S = Y^T$.
When both domains are labeled, then they correspond to different tasks, the transfer task is referred to as inductive learning. 
In this case, the domains may be different or the same, as long as the tasks are different.
If neither domain is labeled and we are trying to establish an understanding or relation between them, that is unsupervised transfer learning. 
Traditional machine learning setup involves both domains being the same and labeled, corresponding to the same task, and does not involve any transfer learning.
Another categorization of transfer learning distinguishes between the features spaces, where homogenous transfer learning is when the source and target domain have the same feature space, and heterogenous transfer learning occurs when the source and target domains have differing feature spaces \cite{day2017survey}.

Within each of these settings of transfer learning there are 4 main approaches to achieve transfer learning, instance-transfer, feature-transfer, parameter transfer, and relational-knowledge transfer \cite{pan2009survey}.
Instance transfer reweighs label data from the source domain in the target domain.
Feature-transfer learns aligned feature representations for the source and target domain.
Parameter transfer shares parameters between the source and target domain models.
Relation-knowledge transfer explicitly models the relation between the source and target domain.

Given the described taxonomies, cross-modal transfer fits as a transductive or heterogeneous transfer learning method. 
Of the 4 approaches to transfer learning, the latter 2 don't make sense to use, thus categorizes works betweenn instance transfer and feature transfer methods for cross modal transfer.
Parameter transfer assumes shared parameters in models from the source and target domains \cite{niu2020decade} and given the drastic difference in domains between modalities, performing only parameter sharing is likely not enough to perform cross-modal transfer and thus no works have shown effective performances solely with this technique. 
Furthermore, mapping precisely between two different modalities is nearly impossible given that they represent information completely differently, and defining a relational knowledge between modalities is less effective and studied than learning a mapping between the datas or reweighing one sensor data type to look like the other, which we categorize as instance based transfer.. 

For the first time, Niu et al. \cite{niu2020decade} introduced the concept of cross-modal learning within the context of a transfer learning survey. 
They claim to be the first transfer learning survey to discuss cross-modal transfer models and argue that most previous transfer methods assume a connection in feature or label spaces, implying that knowledge transfer can only occur when the source and target data are in the same modality. However, cross-modal transfer involves no connection in feature or label spaces; rather, it is based on the abstract semantic meaning behind the data.
Also, previous surveys on heterogeneous transfer learning \cite{day2017survey} have described cross-modal image-text models as part of their transfer learning framework. 
Despite this, whether one subscribes to the data label-based categorization, the cross-modal separate-based categorization, or the input domain difference categorization, it is agreed upon that cross-modality learning can be further categorized into instance-based and feature-based learning. This categorization is presented in this review.

\subsection{Domain Adaptation}
\label{sec::related_works::domain_adaptation}
As stated in \cite{weiss2016survey} there are number of terminology inconsistencies and often times domain adaptation and transfer learning are used interchangeably. 
Here we refer to domain adaptation as the specific technique of altering the source domain to look like the target domain as the means of transferring knowledge to the task \cite{weiss2016survey}.
Arguably, most means of transductive transfer learning involves using domain adaptation.
Domain adaptation allows a machine learning model trained in one domain to efficiently adapt to another related domain for the same output task with fewer data labels \cite{pan2009survey,farahani2021brief}.
Given this focus on scarcely labeled domains, adaptation is often performed through unsupervised \cite{chang2020systematic} or semi-supervised \cite{an2021adaptnet} methods.

In terms of human activity recognition, different data domains can imply adapting between different sensor inputs \cite{bhalla2021imu2doppler}, different positions of wearables on the human body \cite{wang2018deep, chang2020systematic, prabono2021hybrid}, different users \cite{hu2023swl,fu2021personalized} or imu device type \cite{khan2018scaling, zhou2020xhar, chakma2021activity}. 
In the context of this cross-modal learning review, we explore domain adaptation works involving different sensor inputs, which sets us apart from previous HAR based domain adaptation surveys \cite{chang2020systematic}. 



\section{Cross Modal Transfer}
\label{sec::cross_modal_transfer}
Cross-modal transfer was initially introduced in zero-shot learning for image-text modalities, where a deep learning model predicts an object from an unseen class using its knowledge from the text domain\cite{socher2013zero}.
However, multi-modal deep learning for HAR poses a more challenging task as it often uses various time-series devices with different specifications (e.g. device type, frame rates), and as such, it has garnered more recent attention\cite{radu2018multimodal}.
In addition, extrinsic factors such as the location and angle of ambient monitoring sensors or the positioning of wearable device sensors may vary across different settings. 
Traditionally models were trained in an ad hoc fashion where every system needs its own training and testing data, and in a different setting they need to be retrained, but this is time-consuming, computationally inefficient and difficult, particularly with limited data.

Cross-modal learning has emerged as a solution to these difficulties by easily transferring knowledge from one modality to another. 
Some of the resulting benefits of cross-modal learning are enumerated below:
\begin{enumerate}
    \item Each modality might learn better features in each other than if one was trained alone \cite{ngiam2011multimodal}.
    \item It allows one modality to learn a task with little to no training samples if another sensor has learned the same task
    \item It is easy to add or remove sensors to perform the same task allowing for robust modular systems.
    \item Creating strong intermediate representations to transfer between sensors may also help transfer between tasks allowing for easier multi-task generalization of these models
\end{enumerate}

Here we categorize cross-modal learning works into 2 main categories.
Subsection 1 discusses methods which directly map one modality to another, also referred to as instance-based transfer \cite{weiss2016survey}.
Subsection 2 introduces a more generalizable method referred to as feature or representation learning where the model attempts to learn some intermediate representation for the modality and uses that to transfer knowledge across modalities.

\begin{figure}
    \centering
    \includegraphics[width=\columnwidth]{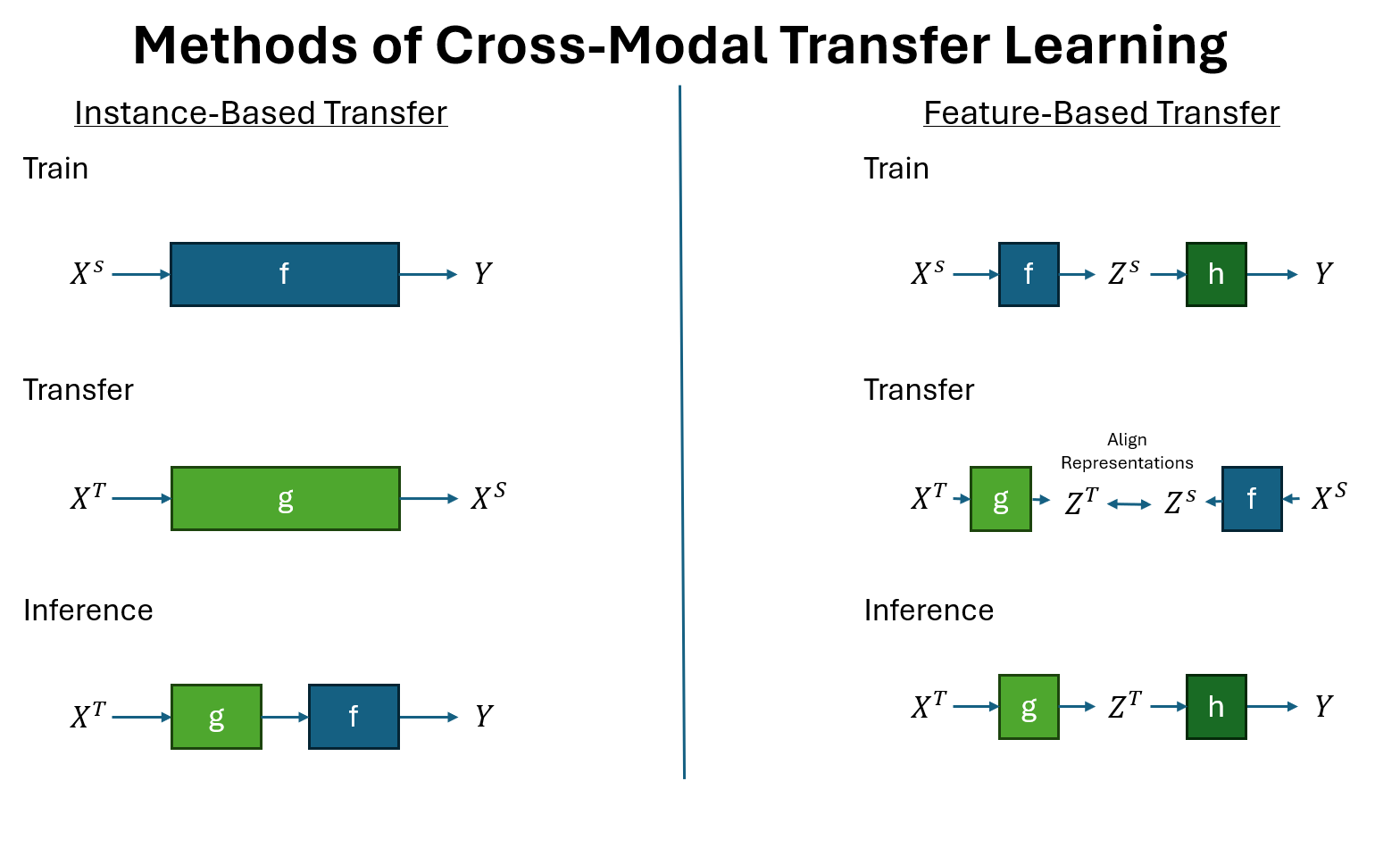}
    \caption{This figure demonstrates the two main methods of cross-modal transfer learning. Instance-based transfer, learns to reweigh the target domain's input $X^T$ as the source domain's input $X^S$ so then inference can be performed using the original trained model $f$. In feature based transfer, the transfer aligns the intermediate representations between the modalities allowing for seamless transfer.
    In cross-modal learning we are only concerned with the case where the input domain spaces are different ($X^S \ne X^S$), thus we have depicted $Y=Y^S = Y^T$}
    \label{fig:cross_modal_learning}
\end{figure}
Assume we have data $x^S$ with label $y^S$ sampled from source dataset $\mathcal{D}^S = {\{(x_i^S,y_i^S)\}}_{i=0}^M, x_i^S \in X^S, y_i^S \in Y^S$.
Suppose we want to generalize the information learned from the source dataset to some target dataset underlying a different distribution $\mathcal{D}^T = {\{(x_i^T,y_i^T)\}}_{i=0}^M, x_i^T \in X^T, y_i^T \in Y^T$. 
We can consider the cases where the inputs are from different domains $X^S \ne X^T$, the case where the outputs are to different tasks $Y^S \ne Y^T$, or the case where both hold true. 
In this literature review, we focus on works that solve the first case of different input domains, specifically different input modalities.

The first method to solve this problem is instance transfer. 
Assume we have trained a model $f:X^S \rightarrow Y$, and we want to train a model $h:X^T \rightarrow Y$ with little or no labeled pairs on the target data $(X^T,Y)$.
Furthermore, we can easily acquire corresponding data between the source and target sensors, i.e. $(X^T,X^S)$ given that collecting this data does not involve human annotation effort.
Instance transfer learns a mapping $g:X^T \rightarrow X^S$ to translate the target sensor data to the source data.
From there, it is easy to transfer the model to the new sensor by simply applying the original model to the translated data $h = f \circ g:X^T \rightarrow Y$.
Alternatively, one can learn the inverse mapping $g^{-1}: X^S \rightarrow X^T$, and use this mapping to curate a dataset $(X^T,Y)$ and then with this data train $h:X^T \rightarrow Y$ without ever needing to manually annotate the target sensor data \cite{banos2021opportunistic}.

The next method that can be leveraged to form a model that can transfer knowledge across input domains is, feature learning, or what is nowadays referred to as multi-modal representation learning.
Representation learning is a method that attempts to learn latent feature vectors $z^S \in Z^S, z^T \in Z^T$ from each domain $X^S, X^T$. 
The idea is that this intermediate latent representation is a highly informative condensed representation of the original modality, for a certain task. 
These information-dense vectors can be used jointly to infer the solution for one task (feature-level sensor fusion), or the two different modalities latent vectors can be used to relate the input domains to transfer knowledge for one or more tasks (feature-based cross-modal learning). 
In this literature review we consider two subcases of feature-based transfer.
The first is where each modality is mapped to a shared representation space.
The second is where each modality is mapped to a related but not necessarily semantically equivalent representation space.

\subsection{Instance-based transfer}
\label{sec::cross_modal_transfer::instance}

Traditionally, direct transfer from one modality to another is performed through whitebox or user-defined solutions that deterministically calculate the translation between sensors requiring prior knowledge of the relation between the domains (e.g. sensor position, data range, units, etc.)
Banos et al. \cite{banos2021opportunistic} take a data-driven black-box approach and learn a network to translate between modalities. 
They argue that white-box solutions are not generalizable or practical, and show promising results on transferring IMU to kinect depth sensor data.
This is one of the first works to leverage deep learning to directly translate between IMU data and another modality in HAR. 

One issue with Banos et al.'s work is that they still had to manually collect corresponding IMU and kinect data to translate between the modalities. 
However, while IMU data is scarce and difficult to label there is an abundance of image, video and text data online, which is one reason unsupervised multimodal models across image, video and text have been so successful. 
Kwon et. al, saw an opportunity to devise a direct transfer method between video and IMU data, enabling the progression of IMU HAR models as more data allows for deeper more complex models. 
Their method, IMUTube \cite{kwon2020imutube, kwon2021complex}, automatically derives accelerometry information from human joints in any given video.
They use a computer vision pipeline that involves 2D pose extraction with a state-of-the-art pose estimator OpenPose, then they lift the 2D keypoints to 3D given some estimates of the camera's extrinsic parameters, and finally they perform post-processing to refine their IMU estimates for the various keypoints. 
Here they use a learned pose estimator as well as some user-defined estimations to perform keypoint lifting and refinement, providing a mixed grey box method compared to Banos et al \cite{banos2021opportunistic}. 
IMUTube has enabled great improvements in video to IMU based cross-modal learning for human activity recognition \cite{kwon2021complex}, as IMUTube allows for deeper more complex models.

CROMOSim \cite{hao2022cromosim} attempts to improve the cross-modality simulation of IMUTube by two main modifications.
First, they use a skinned multiperson linear 3D tri-mesh model as opposed to skeletal keypoints, and they argue this is more realistic as an IMU is placed on the human's skin as opposed to skeletal joints.
This also allows them to simulate IMU positions at any point on the skin as opposed to only on the human joints.
Second, they use a motion capture and video system to train a model that maps video data to skin accelerometry points.
Training a model improves accuracy by eliminating the need to explicitly estimate camera extrincsics (e.g pose and depth), decreasing the computational cost and chance of error.
Nonetheless, the improvements in accuracy trades off with the versatility of the unsupervised IMUTube because although both methods can be applied to un-labled youtube videos CHROMSim requires training on labeled IMU to MoCap data whereas IMUTube does not. 
This illustrates the common feud between two schools of thought for improving machine learning accuracy: a smaller model with less but well-labeled data (CHROMOSim) versus a large model with a high volume of poorly labeled noisy data (IMUTube). 

\subsection{Feature-based Learning}
\label{sec::cross_modal_transfer::feature}

Cross-modal representation learning uses an intermediate representation to help transfer knowledge. 
Here we talk about 2 cases: a) The intermediate reprsentation is the same across all modalities and b) the intermediate representations are different and usually constrained to be somehow related.


\subsubsection{Shared Representation Space}
RecycleML \cite{xing2018enabling} performs different subtasks related to human activity representation by learning some projection for each sensor input into some shared representation space. 
From this joint representation they can train a network to map to the output of the desired task. 
Thus, they can use the last few task-specific layers with a different sensor input and train with fewer datasamples since they only have to re-learn the mapping from sensor to the shared representation space.
As opposed to focusing on learning robust transferable representations, RecycleML focus on HAR related tasks and uses the intermediate representation as a byproduct. 
This approach is practical and effective on limited data in the human activity recognition space, however, would have limited applicability and generalizability to other zero-shot tasks.

\subsubsection{Different Representations}

IMFi \cite{song2021imfi} is a simple example of transferring between modalities using differing representations.
WiFi CSI signal is a unique device free modality for human motion understanding, however, it is often noisy and heavily environment dependent.
IMUs require the user to wear the device however perform fairly consistently independent of the environment.
IMFi learns a mapping from WiFi features to IMU features to perform gait recognition from WiFi consistently in differing environments.
They manually choose IMU features that distinguish gaits and perform drift reduction computation to fine tune them. 
Then similarly perform noise reduction and extract the spectrogram of wifi CSI signals for those corresponding IMU features.
Finally, they learn a CNN classifer that maps the CSI signals to gaits using the IMU features as labels. 
The results showed that transferring knowledge from the IMU features to the CSI classifier allows the CSI gait recognition system to generalize well to unseen environments.

IMU2Doppler \cite{bhalla2021imu2doppler} similarly learns to map IMU representation features in HAR to the less common sensor domain of millimeter wave radar. 
They notice that there exists much more labeled HAR IMU data than for radar datasets; thus, they attempt to transfer knowledge from a learned IMU model to a mmwRadar model.
To do this they first train an IMU model across numerous datasets. Then they simply train the radar data with an additional L2 loss term between the learned representation of the IMU and radar data representation that is being trained: $\mathcal{L}(X_{IMU},X_{radar},Y)= \beta \times |f(X_{IMU})-g(X_{radar})|^2 - \alpha \times \sum_i Y^{(i)} log \hat{Y}^{(i)} $. This forces the representation to be similar to the IMU one, yet not the same.
They show that this method allows them to recognize activities with as little as 15s of labeled radar data.

\cite{tong2021zero} similarly transfers between representations, however, they attempt to work with the more abundant modality video, similar to IMU-Tube.
They claim that training IMU models from IMU data generated from IMUTube still requires many videos and is often dependent on accurate pose tracking to create accurate IMU data, making it sensitive to occlusions and the quality of the video data set. 
To compensate for this \cite{tong2021zero} create a method to directly transfer video representations to IMU representations without needing to simulate the exact accelerometry of the IMU.
This method uses pixel information across the whole video (not just human pose pixels as IMUTube based methods do) and require few video samples to perform well. 
They used a pretrained video model to extract representations of actions in the videos, and then learn a projection from IMU data representation space to this video representation space.
Thus for classes that may be unseen in the IMU dataset, but are seen in the video dataset, they can project the IMU into the video representation space and perform nearest neighbor classification to get a good zero-shot result \cite{tong2021zero}.

One set of related works uses contrastive learning to learn similar representations, as opposed to learning the same representation or different representations and mapping between them.
COCOA \cite{deldari2022cocoa} is one of the first works that investigates sensor based cross-modal contrastive learning.
Previous works have made a lot of progress on contrastive learning as an unsuperivsed method to learn IMU representations \cite{khaertdinov2021contrastive} or to learn features across multiple IMUs \cite{jain2022collossl}, however COCOA takes it a step further and uses contrastive learning across different modalities at the same time instance to align representations.  
Cross modality contrastive learning (COCOA) employs a novel objective function to learn representations from multi-sensor data by computing the cross-correlation between different modalities and minimizing the similarity between different instances.

IMU2CLIP \cite{moon2022imu2clip} focuses on video, IMU, and text data modalities and aligns representations using the Contrastive Language-Image Pretraining (CLIP) method \cite{radford2021learning}.
CLIP is an self-supervised method that uses cosine similarity to score a batch of samples of two different modalities. 
Training updates each modalities encoder such that they generate representations that give a higher score between the same semantic item and lower scores to different items. 
For example, if the text was "a person is jumping" was encoded into a vector and compared to a batch of encoded IMU sequences, the IMU sequence of a person jumping would get a score of 1 in the batch and the other IMU sequences, such as a person sitting, or walking would receive a 0.
Both the text encoder and IMU encoder could then update their respectives weights such that the representation of the text description and IMU sequences align, i.e. have a higher similarity score, and the description with all the negative IMU samples give a lower score.

Although, training brings different modalties'  representations closer together they can never be the same because each modality inherently contains different information.
For example, an IMU sequence of a person running could be better described as "a person running fast" or "a person running uphill" or "a tall person running fast and uphill."
The description can become infinitely more precise thus one IMU sequence should never match one description exactly and thus the representations must also be different.
This method proves fruitful in learning strong representations for various tasks such as motion-based media retrieval, and reasoning tasks \cite{moon2022imu2clip}

ImageBind \cite{girdhar2023imagebind} takes IMU2CLIP's idea a step further and introduces three more modalities, image, thermal and audio.
Instead of training encoders pairwise between each modality, they only train encoders between image/video and the modality, thus using the image modality as the baseline for the representation that binds all these modalities. 
Again, Imagebind shows impressive zero-shot capabilities in cross-modal retrieval, detection, recognition and generation, and using the trained encoder as a feature extractor this method can learn new tasks well with few training samples.
In contrast to RecycleML, IMU2CLIP and ImageBind would perform worse in a specific HAR related tasks, but could likely generalize to other tasks better.

IMU2CLIP and ImageBind differ from COCOA in that they compute only pairwise correlations (dot products) between representations and thus train only two encoders at once, whereas COCOA computes the correlation across all modalities intermediate representations at once, (i.e. the sum of the dot products of each modalities). 
Again here, a CLIP based pairwise training system makes it easy to add a modality as all the modalities encoders wouldn't need to be retrained.
Comparing the performance of the two models is difficult given that IMU2CLIP and ImageBind were trained on egocentric IMU/video pairs whereas COCOA was only trained on multi-imu datasets for HAR and IMU/ECG datasets for emotion recogntion.

\section{Discussion}
\label{sec::discussion}

\subsection{Federated Cross-Modal Learning}

Smart home devices and wearables despite their prevalence, have yet to implement cross-modal transfer in practical applications. 
Seamless deployment is particularly critical when computation is restricted to smaller form factors, and more research should focus on cross modal transfer in resource contained environments.
One related noteworthy solution that has emerged is federated cross-modal learning. 
This approach allows models to be trained collaboratively across multiple devices without centralized data aggregation, catering to the unique challenges presented by Internet of Things (IoT) devices.

Federated learning trains a neural network across multiple devices, often used to address latency, security, and privacy concerns. 
Using multiple modalities in a federated learning framework naturally appears as a multi-modal multi-task problem, as different modalities may appear on different devices but may correspond to the same task. 
Furthermore, different IoT devices are often communicating for different tasks and can naturally provide a partition of the how data is fused or leveraged for different tasks.
These sensors can include ambient monitoring sensors as well as egocentric wearble devices like smart-watches. 
Studying a federated learning setting will likely take us one step closer to seeing an implementation of cross mnodal HAR in smart home or smartcare facilities in the near future.

MM-FedAvg \cite{zhao2022multimodalfederated} introduced multimodal federated learning to improve classification performance and require less client server interaction, as local labeled modalities can be used to train other modalities. 
Specifically they use a semi-supervised learning to learn a global autoencoder across different devices for different modalities.
Another work \cite{xiong2022unified} uses co-attention mechanism to learn complementary information across different modalities and use personalization in federated learning through model-agnostic meta-learning \cite{jiang2019improving,finn2017model} to update the client models and learn the modalities faster based only on local data without having to introduce much extra data.
Contrastive Representation Ensemble and Aggregation for Multimodal FL (CreamFL) \cite{yu2023multimodal} improves further on multimodal representation learning through a global-local cross modal ensemble strategy. 
They use contrastive learning to mitigate discrepancies between modalities and tasks in these ensembles through inter and intra modal negative samples on a public dataset to regularize local training towards a global consensus even for local modality models where the corresponding information is absent. 
Dual Contrastive Multimodal Federate learning (DC-MMFed) \cite{le2023federated} further expands this idea by performing contrastive learning across sensors sharing information locally removing the need for communication to a common public dataset.

\subsection{Integrating Cross-modal transfer and Sensor fusion}
Multi-modal learning is a broad term that encompasses various ways in which modalities in a system can be leveraged together, however, current research focuses on specific methods of using differing modalities.
Works often the sensor based HAR problem often assumes that you need to use a fixed set of sensors and choosing that set from say a phone with so many different sensors is one of the challenges \cite{antar2019challenges}.
Instead, a system should be developed that can dynamically input various number of sensors as data is available and needed given the context. 

Research tends to focus either on transferring knowledge amongst modalities (cross-modal transfer) or on using multiple data types simultaneously (sensor fusion).
Ideally, a system should be able to handle both cases.
For instance, in scenarios where various sensor data are available simultaneously, the system should effectively use all each modality. 
Conversely, when only one type of data is accessible, the system should retain the capacity to transfer relevant knowledge gained from that modalities to the others.

RecycleML successfully transfers knowledge across modalities since each modality shares a common representation space, training one modality for one task seamlessly transfers to another task as it can use the same task specific network. 
Nonetheless, if both modalties are present, Recycle ML doesn't have a method to use both data for the task.

One thought is averaging or somehow combining each modalities representation. 
This then becomes similar to weighted feature fusion sensor fusion methods that could weight a sensor to be zero when the data is not present and learn some reasonable weighting among the sensors given their noise when all the sensors are present. 
HAMLET \cite{islam2020hamlet} is one such method, that uses self-attention to reweigh each modalities features before combing them and passing them through a feedforward network to perform HAR. 
Nonetheless, HAMLET does not map each feature to shared representation space, thus avoiding the benefit of potential cross-modal learning.
Although HAMLET performs well on the UTD MHAD dataset in a multimodal supervised learning task, they claim the degradation on performance in the USCD-MIT dataset maybe due to the absence of sEMG and IMU data in various samples, implying that they could benefit from cross modal learning techniques. 


\subsection{Generalizing to multiple tasks}
In the beginning of this review we gave an overview of many different interelated human motion understanding AI tasks. 
Thus far works usually focus on one of these tasks, or if they focus on multiple tasks they are usually looking at one specific sensor or set number of similar sensors \cite{sheng2020weakly}.
In reality, a system should be able to handle multiple inputs and multiple outputs where each input may or may not be specified and each output may or may not need to specified.

For humans it is so natural to understand a situation through multiple modalities on and off. 
If we're driving a car and we hear a honk we know someone is by us or if we  feel someone hit the car without seeing what happened, or if see and hear someone speeding through a stoplight we might be able to predict that they're going to crash.
Humans seemleslly infer and understand situations with information from only part of the senses, yet most works have a rigid set of assumptions about the data. 
Sensor fusion for multitask HAR usually assumes data is available \cite{sheng2020weakly} and cross-modal learning rarely attempts to infer with multiple devices simultaneously. 

If a multimodal system observe a human walking behind a table or through a door, the IMU should be able to still be able to continue tracking the human.
Yes each modality can do that pretty easily separately, but there is room for the literature to construct joint system that recognizes occlusion or noise in a sensor and adopts or transfers knowledge accordingly.
That too being able to not only localize where the human is and classify what they are doing, but creating a model that can understand and predict human motion is still an open challenge.
All in all, current works fail to explore the range of human motion related tasks and are very narrowly focused.



\subsection{Generative Modeling}
Generative modeling has emerged as a solution to many problems involving "understanding" in AI.
If a model can accurately generate data, it seems to have developed some sort of understanding of that type of data.
The most popular example being large language models (e.g. GPT, Bert) that are simply trained to generate the next token of text.
When trained with vast amounts of data on a large model, this very simple task leads to models that can generalize and perform very complex tasks through prompt engineering.

A similar notion for human motion understanding could be developed from a multi-modal generative model.
Given a sequence of human motions in IMU and camera data, such a model can be used to predict what happens next.
Given multiple sequence of motions over time, the model can be used to forecast future trends and potentially ascribe to a human's health.
As Dall-E can generate images from text, can we generate skeletal data from motion? 
Arguably, our discussion of direct transfer in section \ref{sec::cross_modal_transfer::instance} does imply generating one modality from another such as generating IMU data from video (IMUTube \cite{kwon2020imutube}). 
Again, however, this is very limited to cross-modal transfer task.
Say, for example, both video and IMU data were available, in that case the model would benefit from using both modalities.
Furthermore, learning from simulated data, may produce accuracy problems, as dilineated by subsequent works \cite{hao2022cromosim, tong2021zero}. 
This issue is similar to the Sim2Real gap for simulation based learning methods in robotics. 

Imagebind \cite{girdhar2023imagebind} also showed some data generative capabilities, but they do not test for any other sort of emergent behavior that may imply multi-modal multi-task related abilities in HAR.
In addition, Imagebind was trained with egocentric IMU data from an IMU mounted on a user's head.
Instead, being able to generate a human hands or legs motion trajectory from text may be a more useful task.
Furthermore,  a corresponding video can we localize the humans motion.
One joint generative model could be creatively used in all these situations to solve inumeral HAR related tasks simultaneously.

Unfortunately, training such a generative model not possible given the lack of abundant synchronous modality data as is available online for visual, audio and textual data. 
We could simulate said data, however, we would then have a chicken and egg sort of issue, where we simulate data to train a model to simulate data, and the resulting model would only be as good as the initial simulation calculations we performed. 
Thus research must solve this problem through better models and high quality data as opposed to larger models and lots of noisy data. 

\section{Conclusion}
\label{sec::conclusion}
There are many related cross-modal learning works that could revolutionize AI's ability to understand human motion understanding.
Working both on the implementation aspect of such models through federated learning, the ability to integrate various sensor data into cross modal tasks, generalize to multiple tasks, and abstract to future scenarios are all important future directions of cross-modale learning.

As it stands, cross-modal learning is heavily dominated by feature based methods, also termed as representation learning.
Specifically, the contrastive training paradigm provides promising hopes for learning robust transferrable representations as it is an self-supervised method. 
Nonetheless, this method seems to work best with modalities with abundant data, so it may struggle with data such as IMU, where capturing and psuedo-labeling data is relatively more difficult, than retrieving captioned images and videos from online.

Instance based transfer provides an opportunity to help with this data scarcity problem. 
Dominant methods in this regime, attempt to model IMU data from video data, leveraging existing state of the art pose estimators. 
Although this may help transfer knowledge better from the video to the IMU domain better, the model can only be as good as the original video model, thus negating the potential benefits of learning from different naturally collected modalities.

Overall, human motion understanding is a huge diverse field, and understanding basic activities could be the key to unlocking its potential.
We believe understanding inertial data is the next step in the evolution of AI and learning to effectively transfer between modalities and related tasks is a vital first step. 
In the end, the goal of this paper is to allow researchers to better understand cross-modal learning, to stimulate more research in AI to enable machines to leverage, experience, and interact with multimodal knowledge as humans do to benefit uncountable domains in society from healthcare, entertainments, safety in robotics, social benefits and so on.

{\small
\bibliographystyle{ieeetr}
\bibliography{refs}
}

\end{document}